\RequirePackage{amsmath}
\documentclass[runningheads]{llncs}

\usepackage{multicol}
\usepackage{amsfonts}
					
\usepackage{amssymb}
\usepackage{graphicx}
\usepackage{epic}
\usepackage{eepic}
\usepackage{epsfig,float}
\usepackage{verbatim}
\usepackage{pdfsync}
\usepackage{gastex}
\usepackage{bbm}

\usepackage{xcolor}

\renewcommand{\le}{\leqslant}
\renewcommand{\ge}{\geqslant}

\newcommand{\ol}{\overline}
\newcommand{\eps}{\varepsilon}
\newcommand{\emp}{\emptyset}

\newcommand{\Sig}{\Sigma}

\newcommand{\noin}{\noindent}

\newcommand{\bi}{\begin{itemize}}
\newcommand{\ei}{\end{itemize}}
\newcommand{\be}{\begin{enumerate}}
\newcommand{\ee}{\end{enumerate}}
\newcommand{\bd}{\begin{description}}
\newcommand{\ed}{\end{description}}
\newcommand{\bq}{\begin{quote}}
\newcommand{\eq}{\end{quote}}

\newcommand{\cA}{{\mathcal A}}

\newcommand{\cD}{{\mathcal D}}

\newcommand{\cN}{{\mathcal N}}

\newcommand{\lraL}{{\mathbin{\approx_L}}}

\newcommand{\rev}{\mathbb{R}}
\newcommand{\deter}{\mathbb{D}}
\newcommand{\mini}{\mathbb{M}}

\newcommand{\oc}{$\unlhd$-convex}

\newcommand{\ocl}{$\unlhd$-closed}

\newcommand{\of}{$\unlhd$-free}

\def\shu{\mathbin{\mathchoice
{\rule{.3pt}{1ex}\rule{.3em}{.3pt}\rule{.3pt}{1ex}
\rule{.3em}{.3pt}\rule{.3pt}{1ex}}
{\rule{.3pt}{1ex}\rule{.3em}{.3pt}\rule{.3pt}{1ex}
\rule{.3em}{.3pt}\rule{.3pt}{1ex}}
{\rule{.2pt}{.7ex}\rule{.2em}{.2pt}\rule{.2pt}{.7ex}
\rule{.2em}{.2pt}\rule{.2pt}{.7ex}}
{\rule{.3pt}{1ex}\rule{.3em}{.3pt}\rule{.3pt}{1ex}
\rule{.3em}{.3pt}\rule{.3pt}{1ex}}\mkern2mu}}

\title{Towards a Theory of Complexity of Regular Languages\thanks{This work was supported by the Natural Sciences and Engineering Research Council of Canada 
grant No.~OGP0000871.}
}

\author{Janusz A. Brzozowski\inst{1}}

\authorrunning{J. A. Brzozowski}  

\titlerunning{Complexity of Regular Languages}

\institute{David R. Cheriton School of Computer Science, University of Waterloo \\
Waterloo, ON, Canada N2L 3G1\\
{\tt brzozo@uwaterloo.ca}
}

\begin{document}

%

%
%

\maketitle

\begin{abstract}
We survey recent results concerning the complexity of regular languages represented by their minimal deterministic finite automata.
In addition to the quotient complexity of the language -- which is the number of its (left) quotients, and is the same as its state complexity -- we also consider the size of its syntactic semigroup  and the quotient complexity of its atoms -- basic components of every regular language.
We then turn to the study of the quotient/state complexity of common operations on regular languages: reversal, (Kleene) star, product (concatenation) and boolean operations.
We examine relations among these complexity measures. 
We discuss several subclasses of regular languages defined by  convexity. 
In many, but not all,  cases there exist ``most complex'' languages, languages satisfying all these complexity measures. 
\medskip

\noin
{\bf Keywords:}
atom, boolean operation, complexity measure, concatenation, convex language, most complex language,  quotient complexity, regular language, reversal, star, state complexity,  syntactic semigroup, unrestricted complexity
\end{abstract}

\section{Introduction}
We assume  the reader is familiar with basic properties of regular languages and finite automata, as discussed in~\cite{Per90,Yu97}, for example; formal definitions are given later.

We study the complexity of regular languages represented by their minimal deterministic finite automata (DFAs). 
The number of states in the minimal DFA of a language is its \emph{state complexity}~\cite{Mas70,YZS94}; this number is used as a first measure of complexity.
But languages having the same state complexity can be quite simple or very complex. 
How do we decide whether one language is more complex than another? In this respect, the size of the syntactic semigroup of the language -- which is isomorphic to the transition semigroup of its minimal DFA -- appears to be a good measure.

Another way to distinguish two regular languages of the same state complexity is by comparing how difficult it is to perform operations on these languages.
The state complexity of a regularity preserving unary operation on a language is defined as the maximal complexity of the result of the operation expressed as a function of the state complexity of the language.
For example, we know that there are regular languages of state complexity $n$ whose reverses have state complexity $2^n$, but many languages do not meet this bound.
For binary operations we have two languages of state complexities $m$ and $n$, respectively. The state complexity of a binary operation is the maximal state complexity of the result, expressed as a function of $m$ and $n$.

 In general, to establish the state complexity of a unary operation, we need to find an upper bound on this complexity and a language for each $n$ that meets this bound.
This sequence of languages is called a \emph{stream}.
The languages in the stream often have the same structure and differ only in the parameter $n$.
For binary operations we need two streams.
For some operations the same stream can be used for both operands. 
However, if the second operand cannot be the same as the first, it can usually be a \emph{dialect} of the first operand -- a language that differs only slightly from the first.

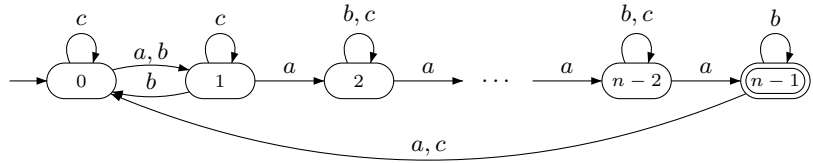
\begin{figure}[ht]
\unitlength 7.5pt
\begin{center}\begin{picture}(37,8)(0,3.5)
\gasset{Nh=1.8,Nw=3.5,Nmr=1.25,ELdist=0.4,loopdiam=1.5}
	{\scriptsize
\node(0)(1,7){0}\imark(0)
\node(1)(8,7){1}
\node(2)(15,7){2}
}
\node[Nframe=n](3dots)(22,7){$\dots$}
	{\scriptsize
\node(n-2)(29,7){$n-2$}
	}
	{\scriptsize
\node(n-1)(36,7){$n-1$}\rmark(n-1)
	}
\drawloop(0){$c$}
\drawedge[curvedepth= .8,ELdist=.1](0,1){$a,b$}
\drawedge[curvedepth= .8,ELdist=-1.2](1,0){$b$}
\drawedge(1,2){$a$}
\drawloop(2){$b,c$}
\drawedge(2,3dots){$a$}
\drawedge(3dots,n-2){$a$}
\drawloop(n-2){$b,c$}
\drawedge(n-2,n-1){$a$}
\drawedge[curvedepth= 4.0,ELdist=-1.0](n-1,0){$a,c$}
\drawloop(n-1){$b$}
\drawloop(1){$c$}
\end{picture}\end{center}
\caption{Minimal DFA of a most complex regular language $L_n(a,b,c)$.}
\label{fig:regular1}
\end{figure}

It has been proved~\cite{Brz13} that the stream $(L_3(a,b,c),\dots,L_n(a,b,c),\dots)$ of  regular languages shown in Fig.~\ref{fig:regular1}  is \emph{most complex} because it meets  the following complexity  bounds: the size of the syntactic semigroup, and the state complexities reversal, (Kleene) star, product/concatenation, and all binary boolean operations. 
It also has the largest number $2^n$ of atoms (discussed later), and all the atoms have maximal state complexity. 

 The \emph{alphabet of a regular language} $L$ is $\Sigma$ (or \emph{$L$ is a language over $\Sigma$}) if $L \subseteq \Sigma^*$ and every letter of $\Sigma$ appears in a word of $L$.
In addition to the usual state complexity of binary operations on languages over the same alphabet,  \emph{unrestricted} state complexity on languages over different alphabets has also been studied~\cite{Brz16}. By adding an input $d$  that induces the identity transformation  in the DFA of Fig.~\ref{fig:regular1}, we obtain a most complex language that
also meets the bounds for unrestricted operations.

A natural question then arises whether most complex language streams also exist in proper subclasses of regular languages. The answer is positive for many, but not all, classes. A rich source of subclasses is provided by the concept of convexity. In this paper we summarize the results for many classes of convex languages. 

 Many of these results were presented as an invited talk at the \emph{20th International Conference on Developments in Language Theory,} Montr\'eal, Qu\'ebec on July 25, 2016. A short abstract appeared in~\cite{Brz16a}.
\section{Quotient/State Complexity of Regular Languages}

Let $\Sig=\{a_1,\dots,a_k\}$ be a nonempty set, called  an  \emph{alphabet}, consisting of \emph{letters} $a_i$, $i=1,\dots, k$. 
A \emph{word} over $\Sig$ is a sequence $a_{i_1}\cdots a_{i_m}$, where $a_{i_j}\in \Sig$, $j=1,\dots,m$; if $m=0$, the word is \emph{empty} and is denoted by $\eps$.
A \emph{language} over $\Sig$ is any subset of $\Sig^*$, where $\Sig^*$ is the free monoid generated by $\Sig$ with $\eps$ as the identity, that is, $\Sig^*$ is the set of all words over $\Sig$.
 Recall that if $L$ is a language over $\Sig$,  every letter of $\Sig$ appears in at least one word of $L$.

The languages $\emp$ (the empty language) and $\{a_i\}$, $i=1,\dots, k$ (the \emph{letter} languages) are called \emph{basic}. A language is \emph{regular} if it can be constructed from the basic languages using only the operations  union (denoted by $L\cup L'$), product (concatenation) (denoted by juxtaposition: $LL'=\{w \mid w=xy, x\in L, y\in L'\}$), and star (denoted by $L^*=
 \bigcup_{n\ge 0} L^n$, where $L^0=\{\eps\}$, and $L^{n+1} = L^nL$).

If $w\in \Sig^*$ and $L\subseteq \Sig^*$, the \emph{(left) quotient} of $L$ by $w$ is the language $w^{-1}L=\{x \mid wx \in L\}$; it is the set of ``all words that can follow $w$ in $L$''.
It is well known that a language is regular if and only if it has a finite number of distinct quotients~\cite{Brz64,Ner58}.
So it is natural to consider the number of quotients of a regular language $L$ as a complexity measure, which we call the \emph{quotient complexity} of $L$ and denote by $\kappa(L)$.  

Quotients  can be computed as follows: For $a,b\in \Sig$, $w\in \Sig^*$ and $L\subseteq \Sig^*$ we have
\begin{eqnarray}
a^{-1}L &=& 
	\left\{ \begin{array}{ll}
		\emp, & \mbox{if $L=\emp$,  or $L=\{b\}$ and $a\not= b$};\\
		\{\eps\}, & \mbox{if $L=a$}.
	\end{array}
	\right.\\
	a^{-1}(L\cup L') &=& a^{-1}L \cup a^{-1}L'.\\
	a^{-1}LL' &=& 
	\left\{ \begin{array}{ll}
		(a^{-1}L)L', & \mbox{if $\eps\notin L$};\\
		(a^{-1}L)L' \cup a^{-1}L', & \mbox{if $\eps \in L$}.
	\end{array}
	\right.\\
	a^{-1}(L^*) &=& (a^{-1}L)L^*.\\
	\eps^{-1}L &=& L.\\
	 (wa)^{-1}L &=& a^{-1}(w^{-1}L).
\end{eqnarray}

When we compute quotients this way, they are represented by expressions involving the basic languages, union, product and star, and it may not be obvious that two different expressions denote the same quotient. However, it is easy to recognize \emph{similarity},  where
two expressions are \emph{similar}~\cite{Brz64} if one can be obtained from the other using the following rules:
\begin{equation}
L\cup L = L, \quad L\cup L'=L'\cup L, \quad L\cup (L'\cup L'')= (L\cup L')\cup L'',
\end{equation}
\begin{equation}
L\cup \emp= L, \quad \emp L= L\emp =\emp, \quad \{\eps\} L= L\{\eps\} =L.
\end{equation}
The number of dissimilar expressions of a regular language is always finite~\cite{Brz64}. 


A concept closely related to a regular language is that of 
a \emph{deterministic finite automaton (DFA)}, which is a quintuple
$\cD=(Q, \Sigma, \delta, q_0,F)$, where
$Q$ is a finite non-empty set of \emph{states},
$\Sig$ is a finite non-empty \emph{alphabet},
$\delta\colon Q\times \Sig\to Q$ is the \emph{transition function},
$q_0\in Q$ is the \emph{initial} state, and
$F\subseteq Q$ is the set of \emph{final} states.
We extend $\delta$ to functions $\delta\colon Q\times \Sig^*\to Q$ and 
$\delta\colon 2^Q\times \Sig^*\to 2^Q$ as usual.
A~DFA $\cD$ \emph{accepts} a word $w \in \Sigma^*$ if ${\delta}(q_0,w)\in F$. The set of all words accepted by $\cD$ is the \emph{language} accepted by $\cD$, denoted by $L(\cD)$. If $q$ is a state of $\cD$, then the language $L_q(D)$ of $q$ is the language accepted by the DFA $(Q,\Sigma,\delta,q,F)$. 
A state is \emph{empty} if its language is empty. Two states $p$ and $q$ of $\cD$ are \emph{equivalent} if 
$L_p(D) = L_q(D)$. 
A state $q$ is \emph{reachable} if there exists $w\in\Sig^*$ such that $\delta(q_0,w)=q$.
A DFA is \emph{minimal} if all of its states are reachable and no two states are equivalent.

The famous theorem of Kleene~\cite{Kle56} states that a language is regular if and only if it is accepted by a DFA. 
We can derive a DFA accepting a regular language $L$ directly from its quotients.
Denote the set of quotients of $L$ by $K=\{K_0,\dots,K_{n-1}\}$, where $K_0=L=\eps^{-1}L$ by convention.
Each quotient $K_i$ can be represented also as $w_i^{-1}L$, where $w_i\in\Sig^*$ is such that
$w_i^{-1}L=K_i$.
Now define the \emph{quotient DFA} of $L$ as follows:
$\cD=(K, \Sig, \delta, K_0, F)$, where $\delta(K_i,a) = K_j$ if $a^{-1}K_i=K_j$, 
and $F=\{K_i\mid \eps \in K_i\}$.
This DFA accepts $L$ and is minimal\footnote{
If a DFA is constructed using dissimilar  expressions and is not minimal,  it can be minimized by one of several methods~\cite{Brz63,Hop71,Moo56}, 
by merging states corresponding to the same expression.}.

In any DFA $\cD=(Q, \Sigma, \delta, q_0,F)$, if $\delta(q_0,w)=q$, then $L_q(D)=L(Q,\Sig,\delta,q,F)$, known also as the \emph{right language of $q$}, is precisely the quotient $w^{-1}L$.
Evidently, the state complexity of a language is equal to its quotient complexity. 
From now on we refer to the quotient/state complexity of $L$ simply as the \emph{complexity} of $L$.

\section{Syntactic/Transition Semigroups}

According to our  complexity measure any two languages with $n$  quotients have the same complexity.
But consider the language $L_n$ accepted by the minimal DFA of Fig.~\ref{fig:regular1} and the language $L'_n=\Sig^{n-2}$. 
Intuitively $L'_n$ is much simpler than $L_n$.

It was proposed in~\cite{Brz13} that the size of the syntactic semigroup of a language should be used as an additional complexity measure.
We proceed to define it now.

The \emph{Myhill congruence} $\lraL$~\cite{Myh57}, also known as the \emph{syntactic congruence},  of a language $L \subseteq \Sig^*$ is defined on $\Sig^+$ as follows:
For $x, y \in \Sig^+$,  
$$x \, \lraL\, y  \text{ if and only if }  wxz\in L  \Leftrightarrow wyz\in L,
 \text{ for all }  w,z \in\Sig^*. $$
The quotient set $\Sig^+/ \lraL$ of equivalence classes of  $\lraL$ is a semigroup, the \emph{syntactic semigroup} $T_{L}$ of $L$.
The \emph{syntactic complexity} of a language $L$ is the cardinality of the syntactic semigroup.

Returning to our example,
the syntactic complexity of $L_n$ is known to be $n^n$, whereas that of $L'_n=\Sig^{n-2}$ is $n-1$; hence syntactic complexity clearly distinguishes the two languages.

 Let $Q_n$ be a set of $n$ elements. Without loss of generality, we assume $Q_n = \{0,1,\dotsc,n-1\}$.
A \emph{transformation} of $Q_n$ is a mapping $t\colon Q_n\to Q_n$.
The \emph{image} of $q\in Q_n$ under $t$ is denoted by $qt$.
If $s,t$ are transformations of $Q_n$, their composition is  defined by
$q(st)=(qs)t$.
Let $\mathcal{T}_{Q_n}$ be the set of all $n^n$ transformations of $Q_n$; then $\mathcal{T}_{Q_n}$ is a monoid under composition. 

For $k\ge 2$, a transformation $t$ of a set $P=\{q_0,q_1,\ldots,q_{k-1}\} \subseteq  Q_n $ is a \emph{$k$-cycle}
if $q_0t=q_1, q_1t=q_2,\ldots,q_{k-2}t=q_{k-1},q_{k-1}t=q_0$.
This $k$-cycle is denoted by  $(q_0,q_1,\ldots,q_{k-1})$, and it acts as the identity on the states not in the cycle.
A~2-cycle $(q_0,q_1)$ is a \emph{transposition}.
A transformation  that sends all the states of $P$ to $q$ and acts as the identity on the remaining states is denoted by $(P \to q)$.  If $P=\{p\}$ we write  $(p\to q)$ for $(\{p\} \to q)$.
 The identity transformation is denoted by $\mathbbm 1$.
 The notation $(_i^j \; q\to q+1)$ denotes a transformation that sends $q$ to $q+1$ for $i\le q\le j$ and is the identity for the remaining states, and  $(_i^j \; q\to q-1)$ is defined similarly.

Let $\mathcal{D} = ( Q_n, \Sigma, \delta, q_0, F)$ be a DFA, where we use $Q_n=\{0,\dots,n-1\}$ as the set of states, without loss of generality.
Each word $w\in \Sigma^+$ induces a transformation $\delta_w$ of the set $Q_n$ defined by $q\delta_w=\delta(q,w)$; we denote this by $w\colon \delta_w$. 
Sometimes we use the word $w$ to denote the transformation it induces; thus we write $qw$ instead of $q\delta_w$.
We  extend the notation to sets of states: if $P\subseteq Q_n$, then $Pw=\{pw\mid p\in P\}$.
We also write $P\stackrel{w}{\longrightarrow} Pw$ to mean that the image of $P$ under $w$ is $Pw$.

The set $T_{\mathcal{D}}$ of all transformations  induced by non-empty words forms a semigroup of transformations called the \emph{transition semigroup} of $\mathcal{D}$~\cite{Pin97}. 
This semigroup is  generated by  $\{\delta_a \mid a \in \Sigma\}$. 
We use the transition semigroup rather than the transition monoid, because the latter always has the identity transformation induced by the empty word, whereas, in the semigroup, if the identity exists it must be induced by a non-empty word. For a more detailed discussion of the necessity of distinguishing between semigroups and monoids see~\cite[Chapter V]{Eil76}, for example.

If  $\cD_n$ is a minimal DFA of $L_n$, then $T_{\cD_n}$ is isomorphic to the syntactic semigroup $T_{L_n}$ of $L_n$~\cite{Pin97}, and we represent elements of $T_{L_n}$ by transformations in~$T_{\cD_n}$. 
We re\-turn to  syntactic complexity later.

\section{Quotients}
Since quotients play a key role in defining a regular language we should also consider their complexity. 
In our example of Fig.~\ref{fig:regular1} all quotients have complexity $n$.
In the case of the language $L'_n=\Sig^{n-2}$, the quotients 
$\eps^{-1}L'_n$, $a^{-1}L'_n$,$\dots$, $a^{n-2}L'_n$, $a^{n-1}L'_n$, where $a\in \Sig$  have complexities $n, n-1,\dots, 2,1$, respectively.
In general, however, the complexity of quotients is not a very good measure because it is always $n$  if the DFA is strongly connected. But to ensure that most complex languages also have most complex quotients, 
we add the complexities of quotients as one of our measures.

\section{Atoms}

Atoms of regular languages were introduced in~\cite{BrTa14} as intersections of quotients.
Atoms as congruence classes were presented in~\cite{Iva16}.
Quotient complexities of atoms were studied in~\cite{BrTa13,Iva16}. 

For a regular language $L$ and words $x,y \in \Sigma^*$ consider the left congruence:
$$x\, {\lhd_L} \,y \mbox{ if and only if } {ux\in L  \Leftrightarrow uy \in L} \mbox { for all } u\in\Sigma^*.$$
An \emph{atom} is a congruence class of $\lhd_L$; thus two words $x$ and $y$ are in the same class if ${x\in u^{-1}L  \Leftrightarrow y \in u^{-1}L} \mbox { for all } u\in\Sigma^*.$
If $Q_n=\{0,\dots,n-1\}$ and $L$ is a regular language with quotients $K = \{K_0,\dotsc,K_{n-1}\}$, 
then each subset $S$ of $Q_n$ defines an \emph{atomic intersection} $A_S = \bigcap_{i \in S} K_i \cap \bigcap_{i \in \ol{S}} \ol{K_i}$, where $\ol{S} = Q_n \setminus S$  and $\ol{L}= \Sig^* \setminus L$ for any $L\subseteq \Sig^*$;
an atom of $L$ is a non-empty atomic intersection. 
It follows that each quotient $K_i$ is a union of atoms, namely of all the atoms in which $K_i$ appears uncomplemented. It is also known that quotients of atoms are unions of atoms~\cite{BrTa14}. Thus atoms are fundamental components of a language, and it was proposed in~\cite{Brz13} that the quotient complexity of atoms should be considered as a complexity measure of regular languages.
 
A \emph{nondeterministic finite automaton (NFA)} is a quintuple
$\mathcal{N}=(Q, \Sigma, \delta, I,F)$, where
$Q$,
$\Sigma$ and $F$ are as in a DFA, 
$\delta\colon Q\times \Sigma\to 2^Q$, and
$I\subseteq Q$ is the \emph{set of initial states}. 
Each triple $(p,a,q)$ with $p,q\in Q$, $a\in\Sig$ is a \emph{transition}  if $q\in \delta(p,a)$.
A sequence $((p_0,a_0,q_0), (p_1,a_1,q_1), \dots, (p_{k-1},a_{k-1},q_{k-1}))$
 of transitions, where $p_{i+1}=q_i$ for $i=0, \dots, k-2$ is a \emph{path} in $\cN$.
The word $a_0a_1\cdots a_{k-1}$ is the word \emph{spelled} by the path. 
A word $w$ is accepted by $\cN$ if there exists a path with $p_0\in I$ and $q_{k-1}\in F$ that spells $w$.

Recall that we have defined the quotient DFA of a regular language $L$ using its quotients as states. In an analogous way, we define an NFA called the \emph{\'atomaton}\footnote{The accent is added to indicate that the word should be pronounced with the stress on the first syllable, and also to avoid confusion between \emph{automaton} and \emph{atomaton}.} of $L$ using atoms as states.
The \'atomaton of $L$ is a NFA $ \cA = ( A, \Sig, \alpha, I_\cA, \{A_{p-1}\} ) $, where
$A$ is the set of atoms of $L$;
$\alpha$ is the transition function defined by $A_j \in \alpha(A_i, a)$ if $aA_j\subseteq A_i$; 
$I_\cA$ is the set of initial atoms, those atoms in which $L=K_0$ appears uncomplemented;
and 
$A_{p-1}$ is the final atom: the only atom containing $\eps$.
In the \'atomaton, the right language of state $A_i$ is the atom $A_i$.

We denote by $L^R$ the reverse  of the language $L$.
Let $\rev$ be the NFA operation that interchanges the sets of initial and final states and reverses all transitions.
Let $\deter$ be the NFA operation that determinizes a given NFA using the subset construction and taking into account only the subsets reachable from the set of initial states.
Finally, let $\mini$ be the minimization operation of DFAs. 
These operations are applied from left to right; thus in $\cN^{\rev\deter\mini\rev}$
the NFA $\cN$ is first reversed, then determinized, then minimized and then reversed again. 

The \'atomaton has the following remarkable properties:

\begin{theorem}[\'Atomaton~\cite{BrTa14}]
\label{thm:atomaton}
Let $L$ be a regular language, let {$\cD$} be its minimal DFA, 
and let {$\cA$} be its \'atomaton. Then 

\be
\item {$\cA$} is isomorphic to {$\cD^{\rev\deter\rev}$}.
\item  {$\cA^\rev$}  is isomorphic to the quotient DFA of $L^R$.
\item {$\cA^\deter$ is isomorphic to $\cD$}.
\item For any NFA $\cN$ accepting $L$, {$\cN^{\rev\deter\mini\rev}$} is isomorphic to  {$\cA$}.
\item {$\cA$} is isomorphic to {$\cD$} if and only if $L$ is {bideterministic}.
\ee
\end{theorem}

A minimal DFA $\cD$ is \emph{bideterministic} if its reverse is also a DFA.
A language is \emph{bideterministic} if its quotient DFA is bideterministic.

The quotient complexity of atoms of  was computed in~\cite{BrTa13} using the \'atomaton. 
To find the complexity of atom $A_i$, the \'atomaton started in state $A_i$ was converted to an equivalent DFA by the subset construction. 
A more direct and simpler method was used in~\cite{Iva16} where  the DFA accepting an atom of a given language is constructed directly from the DFA of the language.

It is clear that any language with $n$ quotients has at most $2^n$ atoms. 
It was proved in~\cite{BrTa13,Iva16} that the following are upper bounds on the 
quotient complexities of atoms:

\begin{equation*}
	\kappa(A_S) \le
	\begin{cases}
		2^n-1, 			& \text{if $S\in \{\emp,Q_n\}$;}\\
		1+ \sum_{x=1}^{|S|}\sum_{y=1}^{n-|S|} \binom{n}{x}\binom{n-x}{y},
		 			& \text{if $\emp \subsetneq S \subsetneq Q_n$.}
		\end{cases}
\end{equation*}

It was shown in~\cite{BrTa13} that the language $L_n$  of Fig.~\ref{fig:regular1} has $2^n$ atoms $A_S$, and each such atom meets the upper bound for the  quotient complexity. 
On the other hand, the language $L'_n=\Sig^{n-2}$ has  atoms:
$\Sig^{n-2},\Sig^{n-3},\dots,\Sig,\eps$. 
Therefore $L'_n$ has only $n-1$ atoms, and its most complex atom has complexity $n$.
Hence atom complexity does distinguish well between $L_n$ and $L'_n$. More will be said about atom complexity later.

The following property of the quotient complexity of atoms was proved by Diekert and Walter~\cite{DiWa15}.
Let  $L_n$ be a language of quotient complexity $n$, and let $f(n)$ be the maximal quotient complexity of its atoms.
Then $f(n+1) / f(n)$ approaches 3 as $n$ approaches infinity.

\section{Quotient Complexity of Operations}

Many software systems have the capability of performing operations on regular languages represented by DFAs.
For such systems it is necessary to know the maximal size of the result of the operation, to have some idea how long the computation will take and how much memory will be required. 
A lower bound on these time and space complexities is provided by the quotient/state complexity of the result of the operation.
For example, suppose we need to reverse a language $L_n$. We apply the reversal operation to a minimal DFA $\cD_n$ of $L_n$ and then use the subset construction to determinize $(\cD_n)^\rev$. 
Since there are at most $2^n$ reachable subsets, we know that $2^n$ is an upper bound on the  state complexity of reversal.
Because we know that this bound can be reached, $2^n$ is a lower bound on the the time and space complexities of reversal.

From now on we denote a language of complexity $n$ by $L_n$, and a DFA with $n$ states, by $\cD_n$.
In general, the \emph{complexity} of a regularity-preserving unary operation $\circ$ on regular languages
is the maximal value of $\kappa(L_n^\circ)$ as a function of $n$, where $L_n$ varies over all regular languages $L_n$ with  complexity $n$.
To show that the bound is tight we need to exhibit a sequence $(L_n, n\ge k)=(L_k,L_{k+1},\dots)$, called a \emph{stream}, of languages that meet this bound. 
The stream does not necessarily start from 1, because the bound may not be reachable for small values of $k$.
In the case of reversal, the stream $(L_3,L_4,\dots)$ of 
Fig.~\ref{fig:regular1} happens to meet the bound for $n\ge 3$. 

In the case of star, Maslov~\cite{Mas70} stated without proof that the tight upper bound for its complexity is $2^{n-1}+2^{n-2}$. A proof was provided by 
Yu, Zhuang and Salomaa~\cite{YZS94}.
This bound is met by the DFA of Fig.~\ref{fig:regular1} for $n\ge 3$.

Next consider the product $L_mL_n$ of two languages $L_m$ and $L_n$. Maslov stated without proof that the tight upper bound for product is $(m-1)2^n+ 2^{n-1}$, and that this bound can be met. 
Yu, Zhuang and Salomaa~\cite{YZS94} showed that there always exists a DFA with at most $(m-1)2^n+ 2^{n-1}$ states that accepts $L_mL_n$, and proved that the bound can be met. 
This bound is also met by $L_m$ and $L_n$ of Fig.~\ref{fig:regular1} for $m,n\ge 3$.

 In general, the \emph{complexity} of a regularity-preserving binary operation $\circ$ on regular languages of complexities $m$ and $n$, respectively, is the maximal value of the result of the operation  as a function of $m$ and $n$, where the operands vary over all regular languages of complexities $m$ and $n$, respectively.
Thus we need two  families $(L'_{m,n}\mid  m \ge h, n\ge k)$ and $( L_{m,n} \mid m \ge h, n \ge k)$ of languages meeting this bound;
the notation $L'_{m,n}$ and $L_{m,n}$ implies that  $L'_{m,n}$ and $L_{m,n}$ depend on both $m$ and $n$. 
Two such examples are known~\cite{HaSa08}: the union and intersection of finite languages require such witnesses. 
However, in all other cases studied in the literature, it is enough to use witness streams $(L'_m, m \ge  h)$ and $(L_n, n\ge k)$, where  $L'_m$ is independent of $n$ and $L_n$ is independent of $m$.

So far we have seen that the stream of Fig.~\ref{fig:regular1} meets the upper bounds for syntactic complexity, quotients, atoms, reversal, star, and product.
The situation is a little different for union (and other binary boolean operations).
Since $L_m\cup L_n$ can have at most $mn$ quotients, we have an upper bound.
Moreover, for $m\neq n$, we know~\cite{Brz13} that the complexity of $L_m\cup L_n$, where these languages are defined in Fig.~\ref{fig:regular1}, does meet the bound $mn$. 
But because $L_n \cup L_n=L_n$, the complexity of union for the languages of 
Fig.~\ref{fig:regular1} is $n$ instead of $n^2$.
So the same stream cannot be used for both arguments.
However, it is possible to use a stream that ``differs only slightly'' from $L_n$
of Fig.~\ref{fig:regular1}.

The notion ``differs only slightly'' is defined as follows~\cite{Brz13,BDL16,BrSi17}.
Let $\Sigma=\{a_1,\dots,a_k\}$ be an alphabet ordered as shown;
if $L\subseteq \Sigma^*$, we denote it by $L(a_1,\dots,a_k)$ to stress its dependence on $\Sigma$.
A \emph{dialect} of $L$ is a  language related to $L$ and obtained by replacing or deleting letters of $\Sigma$ in the words of $L$.
More precisely, for an alphabet $\Sigma'$ and a partial map $\pi \colon \Sigma \mapsto \Sigma'$,
we obtain a dialect of $L$ by replacing each letter $a \in \Sigma$ by $\pi(a)$ in every word of $L$,
or deleting the word entirely if $\pi(a)$ is undefined.
We write $L(\pi(a_1),\dots, \pi(a_k))$ to denote the dialect of $L(a_1,\dots,a_k)$ given by $\pi$,
and we denote undefined values of $\pi$ by  ``$-$''.
For example, if $L(a,b,c)= \{a, ab, ac\}$ then its dialect $L(b,-,d)$ is the language $\{b, bd\}$.
Undefined values for letters at the end of the alphabet are omitted; thus, for example, 
if $\Sigma =\{a,b,c,d,e\}$, $\pi(a)=b$, $\pi(b)=a$, $\pi(c)=c$ and $\pi(d)=\pi(e)=-$, we write $L(b,a,c)$ for $L(b,a,c,-,-)$.

In general, for any binary boolean operation $\circ$ on languages $L_m$ and $L_n$ with quotient DFAs $\cD_m$ and $\cD_n$, to find $L_m\circ L_n$ we use the direct product of 
$\cD_m$ and $\cD_n$ and assign final states in the direct product according to the operation $\circ$. This gives an upper bound of $mn$ for all the operations. 
If we know that the bound $mn$ is met by $L_m \cup L_n$, we also know that the intersection $\ol{L_m} \cap \ol{L_n}$  
meets that bound, because $\kappa(\ol{L})=\kappa(L)$ for all $L$;
similarly, the difference $L_m \setminus \ol{L_n}$ meets that bound. 
It is also known that there are witnesses $L_m$ and $L_n$ such that 
the symmetric difference $L_m \oplus L_n$ meets the bound $mn$. 
A binary boolean function $\circ$ is \emph{proper} if it depends on both of its arguments. 
There are six more proper boolean functions: $\ol{K}\cup\ol{L}=\ol{K\cap L}$, $\ol{K}\cap\ol{L}=\ol{K\cup L}$, $\ol{K}\cup L=\ol{K\setminus L}$, $\ol{K}\cap L=L\setminus K$, $K\cup\ol{L}=\ol{L\setminus K}$, and $\ol{K\oplus L}$.   Thus witnesses for these six functions can be found using the witnesses for union and symmetric difference and their complements.

Our discussion so far, as well as all the literature prior to 2016, used witnesses \emph{restricted} to the same alphabet.
However it is also useful to perform binary operations on languages over different alphabets, for example: $\{ab\} \{ac\}$ or  
$\{a,b\}^*b\cup \{a,c\}^*c$. 
The \emph{unrestricted} complexity of binary operations was first studied in~\cite{Brz16}.
In the case of union and symmetric difference of $L'_m\subseteq (\Sig')^*$ and $L_n \subseteq \Sig^*$, the result is a language over the alphabet $\Sig' \cup \Sig$.
To compute the complexity of $L'_m\cup L_n$, if $L'_m$ does not have an empty quotient, we add an empty state to $\cD'_m$ and send all transitions under letters from $\Sig \setminus \Sig'$ to that state. Similarly, we add an empty state
if needed to $\cD_n$ and send all transitions under letters from $\Sig' \setminus \Sig$ to that state. Thus we have now two languages over the alphabet $\Sig' \cup \Sig$, and we proceed as in the restricted case over the larger alphabet.
It turns out that the complexity of union and symmetric difference is $(m+1)(n+1)$~\cite{Brz16}.

For difference and intersection, $(m+1)(n+1)$ is still an upper bound on their complexity. 
However, the alphabet of $L'_m \setminus L_n$ is $\Sig'$ and the complexity turns out to be $mn+m$ for the difference operation.
Similarly, the alphabet of $L'_m \cap L_n$ is $\Sig' \cap \Sig$, and the complexity of intersection is $mn$, as in the restricted case.
The complexity of any other binary boolean operation can be determined from
the complexities of union, intersection, difference and symmetric difference; 
however, the complexity of $L'_m\circ L_n$ may differ by 1 from 
the complexity of $\ol{L'_m \circ L_n}$.
For more details see~\cite{BrSi16}.

\section{Complexity Measures } 
We have introduced the following measures of complexity for regular languages
$ L_m$ and $L_n$~\cite{Brz13,Brz16}:
\be
\item
The size of the syntactic semigroup of $L_n$.
\item
The complexity of the quotients of $L_n$.
\item
The number of atoms of $L_n$.
\item
The complexity of the atoms of $L_n$.
\item
The complexity of the reverse $L_n^R$ of $L_n$.
\item
The complexity of $L_n^*$, the star of $L_n$.
\item
The restricted and unrestricted complexities of the product $L_mL_n$.
\item
The restricted and unrestricted complexities of boolean operations $L'_m\circ L_n$.
\ee
These measures are not all independent: the relations described below are known.
\begin{theorem}[Semigroup and Reversal  \cite{SWY04}]
\label{prop:SWY}
Let $\cD$ be a minimal DFA with $n$ states accepting a language $L$. If the transition semigroup of $\cD$
has $n^n$ elements, then the complexity of $L^R$ is $2^n$.
\end{theorem}

\begin{theorem}[Number of Atoms and Reversal \cite{BrTa14}]
\label{thm:SWY}
The number of atoms of a regular language $L$ is equal to the complexity of $L^R$.\end{theorem}

Before discussing the next relationships we need to introduce certain concepts from group theory.
If $G$ is a permutation group,  $G$ is \emph{transitive} on a set $X$ if for all $x,y \in X$, there exists $g \in G$ such that $xg = y$.
Also, $G$ is \emph{$k$-set-transitive} if it is transitive on the set of $k$-subsets of $Q_n$, that is, if for all $X,Y \subseteq Q_n$ such that $|X| = |Y| = k$, there exists $g \in G$ such that $Xg = Y$.
If $G$  has degree $n$ and  is $k$-set-transitive for $0 \le k \le n$, then $G$ is  \emph{set-transitive}. 

Set transitive groups have been characterized as follows:

\begin{theorem}[Set Transitive Groups~\cite{BePe55}]
\label{prop:settrans}
A set-transitive permutation group of degree $n$ is {$S_n$} or {$A_n$} or a conjugate of one of the following permutation groups:
\be
\item
For $n=5$, the affine general linear group $\operatorname{AGL}(1,5)$.
\item
For $n=6$, the projective general linear group $\operatorname{PGL}(2,5)$.
\item
For $n=9$, the projective special linear group $\operatorname{PSL}(2,8)$.
\item
For $n=9$, the projective semilinear group $\operatorname{P\Gamma L}(2,8)$.
\ee
\end{theorem}

 We say L is maximally atomic if it has the maximal number of atoms, and each of those atoms has the maximal possible complexity.
 The \emph{rank} of a transformation $t$ is the cardinality of $Q_nt$.
The next result characterizes maximally atomic languages.
\begin{theorem}[Maximally Atomic Languages~\cite{BrDa14a}]
\label{thm:maxatom}
Let $L$ be a regular language over $\Sig$ with complexity $n \ge 3$, and let $T$ be the transition semigroup of the minimal DFA of $L$. Then $L$ is {maximally atomic} if and only if the subgroup of permutations in $T$ is {set-transitive}, and $T$ contains a transformation of {rank $n-1$}.
\end{theorem}

Define the following classes of languages:
\bi
\item
\textbf{FTS} - languages whose minimal DFAs have the {full transformation semigroup} of $n^n$ elements.
\item
\textbf{STS} - languages whose minimal DFAs have transition semigroups with a set-transitive subgroup of permutations and a transformation of rank $n-1$.
\item
 \textbf{MAL} - maximally atomic languages.
 \item
  \textbf{MNA} - languages with the maximal number of atoms.
 \item \textbf{MCR} - languages with a maximally complex reverse.
\ei
The known relations among the various complexity measures are thus as follows:
$$ \text{\textbf{FTS} $\subset$ \textbf{STS} = \textbf{MAL} $\subset$ \textbf{MNA} = \textbf{MCR}} $$

\section{Most Complex Regular Language Streams}
We now exhibit a regular language stream that, together with some dialects, meets  the upper bounds for all complexity measures we have discussed so far~\cite{Brz13,BrSi16}. In this sense this is a \emph{most complex regular language stream} or a \emph{universal witness stream}.
This stream differs from the stream of Fig.~\ref{fig:regular1} only by the identity input $d$.

\begin{definition}
\label{def:regular2}
For $n\ge 3$, let $\cD_n=\cD_n(a,b,c,d)=(Q_n,\Sig,\delta_n, 0, \{n-1\})$, where 
$\Sig=\{a,b,c,d\}$, 
and $\delta_n$ is defined by the transformations
$a\colon (0,\dots,n-1)$,
$b\colon(0,1)$,
$c\colon(n-1 \rightarrow 0)$, and
$d\colon {\mathbbm 1}$.
Let $L_n=L_n(a,b,c,d)$ be the language accepted by~$\cD_n$.
\end{definition}

\begin{theorem}[ Most Complex Regular Languages]
\label{thm:regular2}
For each $n\ge 3$, the DFA of Definition~\ref{def:regular2} is minimal and its 
language $L_n(a,b,c,d)$ has complexity $n$.
The stream $(L_n(a,b,c,d) \mid n \ge 3)$  with some dialect streams
is most complex in the class of regular languages.
In particular, it meets all the complexity bounds below, which are maximal for regular languages.
In several cases the bounds can be met with a reduced alphabet.
\begin{enumerate}
\item
The syntactic semigroup of $L_n(a,b,c)$ has cardinality $n^n$,  and at least three letters are required to meet this bound.  
\item
Each quotient of $L_n(a)$ has complexity $n$.
\item
The reverse of $L_n(a,b,c)$ has complexity $2^n$, and $L_n(a,b,c)$ has $2^n$ atoms.
\item
For each atom $A_S$ of $L_n(a,b,c)$, the complexity $\kappa(A_S)$ satisfies:
 $\kappa(A_S) = 2^n-1,  \text{if $S\in \{\emp,Q_n\}$}$; 
$\kappa(A_S) = 1+ \sum_{x=1}^{|S|}\sum_{y=1}^{n-|S|} \binom{n}{x}\binom{n-x}{y}$, if $\emp \subsetneq S \subsetneq Q_n$.
\item
The star of $L_n(a,b)$ has complexity $2^{n-1}+2^{n-2}$.
\item Product
	\be
	\item Restricted:
	$\kappa( L_m(a,b,c) L_n(a,b,c)) = m2^n-2^{n-1}$.
	\item Unrestricted:
	$\kappa( L_m(a,b,-,c) L_n(b,a,-,d))= m2^n+2^{n-1}$.
	\ee
\item Boolean operations
	\be
	\item
	Restricted:
	For any proper binary boolean operation $\circ$,  
	$\kappa( L_m(a,b) \circ L_n(b,a)) =mn$.
	\item
	Unrestricted:
	$\kappa( L_m(a,b,-,c) \circ L_n(b,a,-,d)) = 
	(m+1)(n+1)$ if $\circ\in \{\cup,\oplus\}$,
	$\kappa( L_m(a,b,-,c) \setminus L_n(b,a)) 
 	= mn+m$, and 
 	$ \kappa( L_m(a,b) \cap L_n(b,a))= mn$. 
 	\ee
 At least four letters are necessary for unrestricted operations{\rm~\cite{Brz16}}.

\end{enumerate}
\end{theorem}
\goodbreak

In the stream above we have used a ``master language'' $L_n$ of Definition~\ref{def:regular2} with four letters, and dialects that use the same alphabet as the master language.
The stream below uses only three letters in the master language of Definition~\ref{def:regular3}, but then adds an extra letter $d$ in a dialect.

\begin{figure}[ht]
\unitlength 8.5pt
\begin{center}\begin{picture}(37,7)(0,3)
\gasset{Nh=1.8,Nw=3.5,Nmr=1.25,ELdist=0.4,loopdiam=1.5}
	{\scriptsize
\node(0)(1,7){0}\imark(0)
\node(1)(8,7){1}
\node(2)(15,7){2}
}
\node[Nframe=n](3dots)(22,7){$\dots$}
	{\scriptsize
\node(n-2)(29,7){$n-2$}
	}
	{\scriptsize
\node(n-1)(36,7){$n-1$}\rmark(n-1)
	}
\drawloop(0){$c$}
\drawedge[curvedepth= .8,ELdist=.2](0,1){$a,b$}
\drawedge[curvedepth= .8,ELdist=-1.2](1,0){$b,c$}
\drawedge(1,2){$a$}
\drawloop(2){$b,c$}
\drawedge(2,3dots){$a$}
\drawedge(3dots,n-2){$a$}
\drawloop(n-2){$b,c$}
\drawedge(n-2,n-1){$a$}
\drawedge[curvedepth= 4.0,ELdist=-1.0](n-1,0){$a$}
\drawloop(n-1){$b,c$}
\end{picture}\end{center}
\caption{Minimal DFA  of  a most complex regular language.}
\label{fig:regular3}
\end{figure}
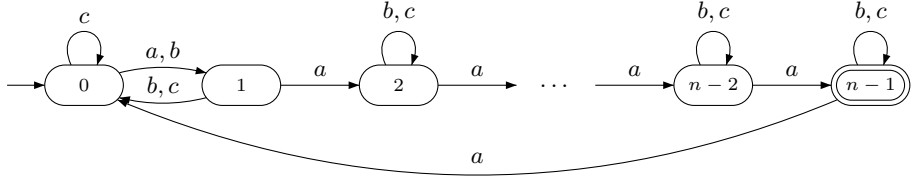

\begin{definition}
\label{def:regular3}
For $n\ge 3$, let $\mathcal{D}_n=\mathcal{D}_n(a,b,c)=(Q_n,\Sigma,\delta_n, 0, \{n-1\})$, where 
$\Sigma=\{a,b,c\}$, 
and $\delta_n$ is defined by the transformations
$a\colon (0,\dots,n-1)$,
$b\colon(0,1)$, and
$c\colon(1 \rightarrow 0)$.
Let $L_n=L_n(a,b,c)$ be the language accepted by~$\mathcal{D}_n$.
The structure of $\mathcal{D}_n(a,b,c)$ is shown in Fig.~\ref{fig:regular3}. 
\end{definition}

\goodbreak

The properties of $L_n$ are the same as those in Theorem~\ref{thm:regular2} except for the following:
\bi
\item
The bound for the restricted product is met by $L_m(a,b) L_n(a,-,b)$.
\item
The bound for the unrestricted product is met by $L_m(a,b) L_n(a,c,b)$.
\item
The bound for the unrestricted union and symmetric difference is met by $L_m(a,b,c) L_n(b,a,d)$.
\item
The bound for the unrestricted difference is met by $L_m(a,b,c) L_n(b,a)$.
\ei
\medskip

The most complex streams introduced in this section will be used in several subclasses of regular languages.

\section{Most Complex Languages in Subclasses}

Many interesting proper subclasses of the class of regular languages can be defined using the notion of convexity.
Convex languages were introduced in 1973 by Thierrin~\cite{Thi73} and revisited in 2009 by Ang and Brzozowski~\cite{AnBr09}.

Convexity can be defined with respect to any binary relation on $\Sig^*$.
Let $\unlhd$ be such a binary relation; if $u\unlhd v$ and 
$u\not = v$, we write $u\lhd v$.
Let $\unrhd$ be the converse binary relation, that is, let $u\unrhd v$ if and only if $v \unlhd u$.
A~language $L$ is {\em \oc{}\/}
if $u\unlhd v$,  $u\unlhd w$, 
and $v\unlhd w$ with $u,w\in L$ imply $v\in L$.
It is {\em \of{}\/}  if $v\lhd w$ and   $w\in L$ imply $v\not\in L$.
It is {\em \ocl{}\/}  if $v\unlhd w$ and   $w\in L$ imply $v\in L$.
It is {\em $\unrhd$-closed\/}  if $v\unrhd w$ and   $w\in L$ imply $v\in L$.
Languages that are $\unrhd$-closed are also called \emph{$\unlhd$-converse-closed}.
One verifies that a language is $\unlhd$-closed if and only if its complement is $\unrhd$-closed.

If $w = xyz$, where $x, y, z \in \Sig^*$, then $x$ is a \emph{prefix} of $w$, $y$ is a \emph{factor} of $w$, and $z$ is a \emph{suffix} of $w$. Note that a prefix or a suffix is also a factor.
If $w=w_0a_1w_1\cdots a_nw_n$, where $a_1,\ldots,a_n\in \Sigma$, and $w_0,\ldots,w_n\in\Sigma^*$, then $v=a_1\cdots a_n$ is a {\em subword\/} of $w$; note that every factor of $w$ is a subword\footnote{The word ``subword'' is often used to mean ``factor''; here by a ``subword''  we mean a subsequence.} of $w$.

The \emph{shuffle} $u\shu v$ of words $u,v\in\Sig^*$ is defined as follows:
$$ u\shu v= \{u_1v_1\cdots u_kv_k \mid u=u_1\cdots u_k,  v=v_1\cdots v_k,
u_1,\ldots,u_k,v_1,\ldots, v_k\in \Sig^*\}.$$
The shuffle of two languages $K$ and $L$ over $\Sig$ is defined by
$$ K\shu L=\bigcup_{u \in K, v \in L} u\shu v.$$
Note that the shuffle operation is commutative on both words and languages.

Here we consider only four binary relations for defining convexity: ``is a prefix of'', ``is a suffix of'',  ``is a factor of'', and ``is a subword of''. 
Each of these four relations is a partial order on $\Sig^*$ and leads  to
four classes of languages; we illustrate this using the prefix relation:
\bi
\item
A language $L$ that is \emph{prefix-converse-closed} is a \emph{right ideal}, that is, it satisfies the equation  $L=L\Sig^*$.
\item
A language $L$ that is \emph{prefix-closed} is the complement of a right ideal.
\item
A language that is \emph{prefix-free} and not $\{\eps\}$ is a prefix-code~\cite{BPR10}.
\item
A language  is \emph{proper prefix-convex} if it not a right ideal and is neither closed nor free.
\ei

Similarly, we define \emph{suffix-converse-closed} languages which are \emph{left ideals} (satisfy $L=\Sig^* L$), \emph{suffix-closed}, \emph{suffix-free} (\emph{suffix codes}~\cite{BPR10}),
and \emph{proper suffix-convex} languages, 
\emph{two-sided ideals} 
(that satisfy $L=\Sig^* L \Sig^*$), 
\emph{factor-closed}, \emph{factor-free} (\emph{infix codes}~\cite{ShTh74}), and \emph{proper factor-convex} languages, and also 
\emph{subword-converse-closed} languages which are 
\emph{all-sided ideals} (that satisfy $L=L\shu \Sig^*$),
\emph{subword-closed},
\emph{subword-free} (\emph{hypercodes}~\cite{ShTh74}), and 
\emph{proper subword-convex} languages.

Decision problems for convex languages were studied in~\cite{BSX11}.
We can decide in $O(n^3)$ time if a given regular language $L$  over a fixed alphabet $\Sig$  accepted by a DFA with $n$ states is prefix-, suffix-, factor-, and subword-convex.
We can decide in $O(n^2)$ time if $L$  is 
prefix-free,
left ideal,
suffix-closed,
suffix-free,
two-sided ideal,
factor-closed, 
factor-free,
all-sided ideal,
subword-closed,
subword-free.
We can decide in $O(n)$ time if $L$  is a
right ideal or a
prefix-closed language.
\smallskip

We now consider the complexity properties of some convex languages.

\subsection{Prefix-Convex Languages}

{\bf RIGHT IDEALS} The complexity of right ideals was studied as follows:
complexities of common operations using various witnesses~\cite{BJL13},
semigroup size~\cite{BrYe11}, complexities of atoms~\cite{BrDa15}, 
most complex right ideals with restricted operations~\cite{BDL16},
most complex right ideals with restricted and unrestricted operations and four-letter witnesses~\cite{BrSi17},
most complex right ideals with restricted and unrestricted operations and five-letter witnesses~\cite{BrSi16}.
Here we use the witnesses from~\cite{BrSi17}.

\begin{definition}
\label{def:ideal}
For $n\ge 4$, let $\mathcal D_n=\mathcal D_n(a,b,c,d)=(Q_n,\Sigma,\delta_n, 0, \{n-1\})$, where
$\Sigma=\{a,b,c,d\}$
and $\delta_n$ is defined by 
$a\colon (0,\dots,n-2)$,
$b\colon(0,1)$,
$c\colon(1 \rightarrow 0)$, and
$d\colon (_0^{n-2} q\to q+1)$.
This DFA uses the structure of Fig.~\ref{fig:regular3} for the states in $Q_{n-1}=\{0, \dots, n-2\}$ and letters in $\{a,b,c\}$.
Let $L_n=L_n(a,b,c,d)$ be the language of~$\mathcal D_n$.
\end{definition}

\begin{theorem}[Most Complex Right Ideals]
\label{thm:rightideals}
For each $n\ge 4$, the DFA of Definition~\ref{def:ideal} is minimal and  $L_n(a,b,c,d)$ is a right ideal of  complexity $n$.
The stream $(L_n(a,b,c,d) \mid n \ge 4)$  with some dialect streams 
is most complex in the class of  right ideals.
It meets  the following bounds:
1. Semigroup size: $n^{n-1}$. 
2.  Quotient complexities: $n$, except $\kappa(\Sig^*)=1$.
3. Reversal: $2^{n-1}$. 
4. Atom complexities: 
 $\kappa(A_S) = 2^{n-1},  \text{if $S=Q_n$}$; 
$\kappa(A_S) = 1 + \sum_{x=1}^{|S|}\sum_{y=1}^{n-|S|}\binom{n-1}{x-1}\binom{n-x}{y}$, if $\emptyset \subsetneq S \subsetneq Q_n$.
5. Star: $n+1$. 
6. (a) Restricted product: $m+2^{n-2}$;
(b) Unrestricted  product: $m+2^{n-1}+2^{n-2}+1$.
7. (a)  Restricted boolean operations:
 $mn$ if $\circ\in \{\cap,\oplus\}$, 
 $mn-(m-1)$ if $\circ=\setminus$, and $mn-(m+n-2)$ if $\circ=\cup$.
(b) Unrestricted boolean operations: same as regular languages.
 At least four letters are required to meet all these bounds{\rm ~\cite{BSY15}}.
\label{thm:ideal}
\end{theorem}

\noin
{\bf PREFIX-CLOSED LANGUAGES} The complexities of common operations on prefix-closed languages using various witnesses were studied in~\cite{BJZ14,HSW09}. Most complex prefix-closed languages were examined in~\cite{BrSi17}. As every prefix-closed language has an empty quotient, the restricted and unrestricted complexities are the same.

\begin{definition}
\label{def:ClosedWit}
For $n\ge 4$, let $\mathcal{D}_n=\mathcal{D}_n(a,b,c,d)=(Q_n,\Sigma,\delta_n, 0, Q_n\setminus \{n-1\})$, 
where 
$\Sigma=\{a,b,c,d\}$, 
and $\delta_n$ is defined by $a\colon (0,\dots,n-2)$,
$b\colon(0,1)$,
${c\colon(1\rightarrow 0)}$, and
$d\colon \left(_{n-2}^{0} \; q\to q-1 \pmod n \right)$.
Let $L_n=L_n(a,b,c,d)$ be the language of~$\mathcal{D}_n$.
\end{definition}

\begin{theorem}[Most Complex Prefix-Closed Languages]
\label{thm:closed}
 For $n\ge 4$, the DFA of Definition~\ref{def:ClosedWit} is minimal and  $L_n(\Sigma_n)$ is a prefix-closed language of complexity $n$.
The stream $(L_m(a,b,c,d) \mid m \ge 4)$ with some dialect streams
is most complex in the class of prefix-closed languages, and meets the following bounds:
1. Semigroup size: $n^{n-1}$. 
2.  Quotient complexities: $n$, except $\kappa(\emp)=1$.
3. Reversal: $2^{n-1}$. 
4. Atom complexities:
 $\kappa(A_S) = 2^{n-1}$, if $S=\emptyset$; 
$\kappa(A_S) =1 + \sum_{x=1}^{n-|S|}\sum_{y=1}^{|S|}\binom{n-1}{x-1}\binom{n-x}{y}$, if $\emptyset \subsetneq S \subsetneq Q_n$.
5. Star $2^{n-2}+1$. 
6. Product: $(m+1)2^{n-2}$.
7. Boolean operations: $mn$ if $\circ\in \{\cup,\oplus\}$, 
 $mn-(n-1)$ if $\circ=\setminus$, and $mn-(m+n-2)$ if $\circ=\cap$. 
  At least four letters are required to meet all these bounds~{\rm \cite{BrSi17}}.
\end{theorem}

\noin
{\bf PREFIX-FREE LANGUAGES} The complexities of operations on prefix-free languages with various witnesses were studied in~\cite{HSW09,JiKr10,Kra11}.
 The syntactic complexity bound of $n^{n-2}$ was established in~\cite{BLY12}.
 Most complex prefix-free languages were  considered in~\cite{BrSi17}.
As every prefix-free language has an empty quotient, the restricted and unrestricted complexities are the same for binary operations. 

\begin{definition}
\label{def:prefix-free}
For $n\ge 4$, 
let $\Sigma_n=\{a,b,c,d,e_0,\dots,e_{n-3}\}$ and let DFA $\mathcal{D}_n(\Sig_n)$ be
$\mathcal{D}_n(\Sigma_n) = (Q_n,\Sigma_n,$
$\delta_n,0,\{n-2\}),$ 
where  
$\delta_n$ is defined by 
$ a \colon (n-2 \to n-1) (0,\dots,n-3) $,
$ b \colon (n-2 \to n-1) (0,1) $, 
$ c \colon (n-2 \to n-1) (1\to 0)  $,
$ d \colon (0\to n-2) (Q_n\setminus \{0\} \to n-1)$,
$e_q\colon (n-2\to n-1)(q \to n-2)$ for $q=0,\dots,n-3$.
The transformations induced by $a$ and $b$ coincide when $n=4$.
This DFA uses the structure of the DFA of Fig.~\ref{fig:regular3} for the states in $Q_{n-2}=\{0, \dots, n-3\}$ and letters in $\{a,b,c\}$.
Let $L_n(\Sigma_n)$ be the language of $\mathcal{D}_n(\Sigma_n)$.
\end{definition}

\begin{theorem}[Most Complex Prefix-Free Languages]
\label{thm:Prefix-free_witness}
For $n\ge 4$, the DFA of Definition~\ref{def:prefix-free} is minimal and  $L_n(\Sigma_n)$ is a prefix-free language of complexity $n$.
The stream $(L_n(a,b,c,d,e_0,\dots,e_{n-3}) \mid n \ge 4)$  with some dialect streams
is a most complex prefix-free language.
At least $n+2$ inputs are required to meet all the bounds below~\cite{BrSi17}:
1. Semigroup size: $n^{n-2}$. 
2.  Quotient complexities: $n$, except $\kappa(\eps)=2$, $\kappa(\emp)=1$.
3. Reversal: $2^{n-2}+1$.
4. Atom complexities:
 $\kappa(A_S) = 2$, if $S=\{n-2\}$; 
$\kappa(A_S) = 2^{n-1}$, if $S= \emptyset$;
$\kappa(A_S) = 2^{n-2}+1$, if $S= Q_{n-2}$; 
$\kappa(A_S) =  2 + \sum_{x=1}^{|S|}\sum_{y=1}^{n-2-|S|}\binom{n-2}{x}\binom{n-2-x}{y}$, if $\emptyset \subsetneq S \subsetneq Q_{n-2}$.
5. Star: $n$.
6. Product: $m+n-2$.
7. Boolean operations: $mn-2$ if $\circ\in \{\cup,\oplus\}$, 
 $mn-(m+2n-4)$ if $\circ=\setminus$, and $mn-2(m+n-3)$ if $\circ=\cap$. 

\end{theorem}

\begin{figure}[th]
\unitlength 7pt
\begin{center}\begin{picture}(40,14)(0,-1.6)
\gasset{Nh=2.2,Nw=6.0,Nmr=1.25,ELdist=0.4,loopdiam=1.5}
{\footnotesize
\node(0)(-1,7){0}\imark(0)
\node(1)(9.5,7){1}
\node(2)(18,7){2}
\node[Nframe=n](3dots)(26.5,7){$\dots$}

\node(n-2-k)(35,7){$n-2-k$}

\node(n-1)(41,3.5){$n-1$}

\node(n-1-k)(-1,0){$n-1-k$}\rmark(n-1-k)
		\node(n-k)(9.5,0){$n-k$}\rmark(n-k)

\node(n-k+1)(18,0){$n-k+1$}\rmark(n-k+1)
	
\node[Nframe=n](3dots1)(26.5,0){$\dots$}
\node(n-2)(35,0){$n-2$}\rmark(n-2)
\drawedge[curvedepth= 1.2,ELdist=-1.3](0,1){$b, d_1$}
\drawedge(n-2-k,n-1){$d_1$}
\drawedge(0,n-1-k){$e$}
\drawedge[curvedepth= 1.2,ELdist=.2](1,0){$b,c_1$}
\drawedge[curvedepth= -4.5,ELdist=.3](n-2-k,1){$a$}
\drawedge(1,2){$a, d_1$}
\drawedge(2,3dots){$a, d_1$}
\drawedge(3dots,n-2-k){$a, d_1$}
\drawedge[curvedepth= 1.2,ELdist=-1.3](n-1-k,n-k){$a, b, d_2$}
\drawedge[curvedepth= 1.2,ELdist=.25](n-k,n-1-k){$a,c_2$}
\drawedge(n-k,n-k+1){$b, d_2$}
\drawedge(n-k+1,3dots1){$b, d_2$}
\drawedge(3dots1,n-2){$b, d_2$}
\drawedge(n-2,n-1){$d_2$}
\drawedge[curvedepth= -4.5,ELdist=.3](n-2,n-1-k){$b$}
}
\end{picture}\end{center}
\caption{DFA $\mathcal{D}_{n,k}(a,b,c_1,c_2, d_1, d_2,e)$ of Definition~\ref{def:proper} when $n-1-k$ is even, $k$ is odd, and both are at least $2$; missing transitions are self-loops.}
\label{fig:ProperWit2}
\end{figure}
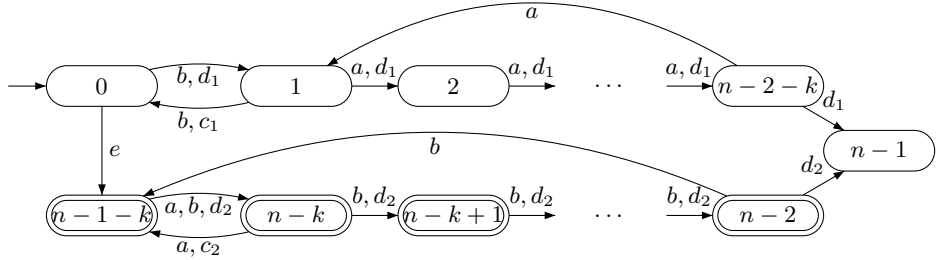

\noin
{\bf PROPER PREFIX- CONVEX LANGUAGES}
Proper prefix-convex languages were studied in~\cite{BrSi16a}. 
In contrast to the three special cases, they represent the full nature of prefix-convexity.
\begin{definition}\label{def:proper}
For $n \ge 3$,  $1 \le k \le n-2$,  let $\mathcal D_{n,k}(\Sigma) = (Q_n, \Sigma, \delta_{n,k}, 0, F_{n,k})$ where $\Sigma = \{a, b, c_1, c_2, d_1,d_2,e\}$, $F_{n,k} = \{n-1-k, \dots , n-2\}$, and $\delta_{n,k}$ is given by  the transformations below.

Also, let $E_{n,k} = \{0, \dots, n-2-k\}$; it is useful to partition $Q_n$ into $E_{n,k}$, $F_{n,k}$, and $\{n-1\}$.
Letters $a$ and $b$ have complementary behaviours on $E_{n,k}$ and $F_{n,k}$, depending on the parities of $n$ and $k$.
Letters $c_1$ and $d_1$ act on $E_{n,k}$  exactly in the same way as $c_2$, and $d_2$ act on $F_{n,k}$.
In addition, $d_1$ and $d_2$ send states $n-2-k$ and $n-2$, respectively, to state $n-1$, and letter $e$ connects the two parts of the DFA.
The structure of $\mathcal{D}_n(\Sigma)$ is shown in Figs.~\ref{fig:ProperWit1} and \ref{fig:ProperWit2} for certain parities of $n-1-k$ and $k$. 
Let $L_{n,k}(\Sigma)$ be the language recognized by $\mathcal D_{n,k}(\Sigma)$.

\begin{align*}
a &\colon \begin{cases}
(1, \dots, n-2-k)(n-1-k, n-k), &\text{ \emph{if} $n-1-k$ \emph{is even and} $k \ge 2$;} \\
(0, \dots, n-2-k)(n-1-k, n-k), &\text{ \emph{if} $n-1-k$ \emph{is odd and} $k \ge 2$;} \\
(1, \dots, n-2-k), &\text{ \emph{if} $n-1-k$ \emph{is even and} $k = 1$;} \\
(0, \dots, n-2-k), &\text{ \emph{if} $n-1-k$ \emph{is odd and} $k = 1$.} \\
\end{cases}\\
 b &\colon \begin{cases}
(n-k, \dots, n-2)(0, 1), &\text{ \emph{if} $k$ \emph{is even and} $n-1-k \ge 2$;} \\
(n-1-k, \dots, n-2)(0, 1), &\text{ \emph{if} $k$ \emph{is odd and} $n-1-k \ge 2$;} \\
(n-k, \dots, n-2), &\text{ \emph{if} $k$ \emph{is even and} $n-1-k = 1$;} \\
(n-1-k, \dots, n-2), &\text{ \emph{if} $k$ \emph{is odd and} $n-1-k = 1$.} \\
\end{cases}\\
c_1 &\colon \begin{cases}(1 \to 0), &\text{\emph{if} $n-1-k \ge 2$;} \\ \quad~ \mathbbm{1}, &\text{\emph{if} $n-1-k = 1$.} \end{cases}\\
c_2 &  \colon \begin{cases}(n-k \to n-1-k), &\text{\emph{if} $k \ge 2$;} \\ \quad\quad\quad~ \mathbbm{1}, &\text{\emph{if} $k = 1$.} 
\end{cases}\\
d_1 &\colon (n-2-k \rightarrow n-1)(_0^{n-3-k} \;\; q\rightarrow q+1).\\
d_2 &\colon (_{n-1-k}^{n-2} \;\; q\rightarrow q+1).\\
e &\colon (0 \to n-1-k).
\end{align*}

\end{definition}

\begin{figure}[th]
\unitlength 7pt
\begin{center}\begin{picture}(40,14)(0,-1.6)
\gasset{Nh=2.2,Nw=6.0,Nmr=1.25,ELdist=0.4,loopdiam=1.5}
{\footnotesize
\node(0)(-1,7){0}\imark(0)
\node(1)(9.5,7){1}
\node(2)(18,7){2}
\node[Nframe=n](3dots)(26.5,7){$\dots$}
	
\node(n-2-k)(35,7){$n-2-k$}

\node(n-1)(41,3.5){$n-1$}

\node(n-1-k)(-1,0){$n-1-k$}\rmark(n-1-k)

\node(n-k)(9.5,0){$n-k$}\rmark(n-k)

\node(n-k+1)(18,0){$n-k+1$}\rmark(n-k+1)
	
\node[Nframe=n](3dots1)(26.5,0){$\dots$}
\node(n-2)(35,0){$n-2$}\rmark(n-2)
\drawedge[curvedepth= 1.2,ELdist=-1.3](0,1){$a, b, d_1$}
\drawedge(n-2-k,n-1){$d_1$}
\drawedge(0,n-1-k){$e$}
\drawedge[curvedepth= 1.2,ELdist=.2](1,0){$b,c_1$}
\drawedge[curvedepth= -4.5,ELdist=.3](n-2-k,0){$a$}
\drawedge(1,2){$a, d_1$}
\drawedge(2,3dots){$a, d_1$}
\drawedge(3dots,n-2-k){$a, d_1$}
\drawedge[curvedepth= 1.2,ELdist=-1.3](n-1-k,n-k){$a, d_2$}
\drawedge[curvedepth= 1.2,ELdist=.25](n-k,n-1-k){$a,c_2$}
\drawedge(n-k,n-k+1){$b, d_2$}
\drawedge(n-k+1,3dots1){$b, d_2$}
\drawedge(3dots1,n-2){$b, d_2$}
\drawedge(n-2,n-1){$d_2$}
\drawedge[curvedepth= -4.5,ELdist=.3](n-2,n-k){$b$}
}
\end{picture}\end{center}
\caption{DFA $\mathcal{D}_{n,k}(a,b,c_1,c_2, d_1, d_2,e)$ of Definition~\ref{def:proper} when $n-1-k$ is odd, $k$ is even, and both are at least $2$; missing transitions are self-loops.}
\label{fig:ProperWit1}
\end{figure}
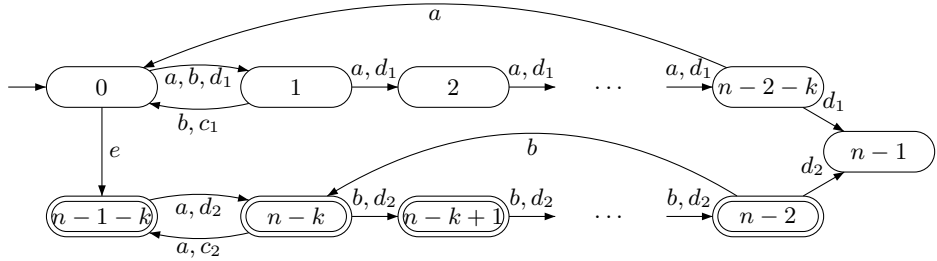

\begin{theorem}[Proper Prefix-Convex Languages]
\label{thm:propermain}
For $n\ge 3$ and $1 \le k \le n-2$, the DFA $\mathcal{D}_{n,k}(\Sigma)$ of Definition~\ref{def:proper} is minimal and  $L_{n,k}(\Sigma)$ is a $k$-proper language of complexity $n$.  
The bounds below are maximal for $k$-proper prefix-convex languages. At least seven letters are required to meet these bounds.
\begin{enumerate}
\item
The syntactic semigroup of $L_{n,k}(\Sigma)$ has cardinality $n^{n-1-k}(k+1)^k$; the maximal value $n(n-1)^{n-2}$ is reached only when $k=n-2$.
\item
The non-empty, non-final quotients of $L_{n,k}(a, b, -, -, -, d_2, e)$ have complexity $n$, the final quotients have complexity $k+1$, and $\emp$
has complexity 1.
\item
The reverse of $L_{n,k}(a,b,-,-,-,d_2,e)$ has complexity $2^{n-1}$; moreover, the language $L_{n,k}(a,b,-,-,-,d_2,e)$ has $2^{n-1}$ atoms for all $k$.
\item 

For each atom $A_S$ of $L_{n,k}(\Sigma)$, write $S = X_1 \cup X_2$, where $X_1 \subseteq E_{n,k}$ and $X_2 \subseteq F_{n,k}$.
Let $\overline{X_1} = E_{n,k}\setminus X_1$ and $\overline{X_2} = F_{n,k}\setminus X_2$.
If $X_2 \not= \emptyset$, then 
{\small
 $\kappa(A_S) =$
$1 + \sum_{x_1=0}^{|X_1|}\sum_{x_2=1}^{|X_1| + |X_2| - x_1}\sum_{y_1=0}^{|\overline{X_1}|}\sum_{y_2=0}^{|\overline{X_1}|+|\overline{X_2}| -y_1}\binom{n-1-k}{x_1}\binom{k}{x_2}\binom{n-1-k-x_1}{y_1}\binom{k-x_2}{y_2}.$
}
If $X_1 \not= \emptyset$ and $X_2 = \emptyset$, then 
{\footnotesize 
$\kappa(A_S) = 1+$\\
$ \sum_{x_1=0}^{|X_1|}\sum_{x_2=0}^{|X_1| - x_1}\sum_{y_1=0}^{|\overline{X_1}|}\sum_{y_2=0}^{k}\binom{n-1-k}{x_1}\binom{k}{x_2}\binom{n-1-k-x_1}{y_1}\binom{k-x_2}{y_2} 
-2^k\sum_{y=0}^{|\overline{X_1}|}\binom{n-1-k}{y}.$
}
Otherwise, $S = \emptyset$ and $\kappa(A_S) = 2^{n-1}$.

\item
The star of $L_{n,k}(a,b,-,-,d_1,d_2, e)$  has complexity $2^{n-2}+2^{n-2-k}+1$. The maximal value $2^{n-2}+2^{n-3}+1$ is reached only when $k=1$.
\item
$ L_{m,j}(a,b,c_1,-, d_1, d_2, e) L_{n,k}(a,d_2, c_1,-,d_1, b, e)$ has complexity $m-1-j +j2^{n-2}+2^{n-1}$. The maximal value $m 2^{n-2} + 1$ is reached only when $j=m-2$.
\item
For $m,n\ge 3$, $1 \le j \le m-2$, and $1 \le k \le n-2$, define the languages $ L_{m,j} =  L_{m,j}(a, b, c_1, -, d_1, d_2, e)$ and $L_{n,k} = L_{n,k}(a, b, e, -, d_2, d_1, c_1)$. For any proper binary boolean function $\circ$, the complexity of $ L_{m,j} \circ L_{n,k}$ is maximal. Thus
	\begin{enumerate}
	\item
	$ L_{m,j} \cup L_{n,k}$ and $ L_{m,j} \oplus L_{n,k}$ have complexity  $mn$.
	\item
	$ L_{m,j} \setminus L_{n,k}$ has complexity $mn-(n-1)$.
	\item
	$ L_{m,j} \cap L_{n,k}$ has complexity $mn-(m+n-2)$.
	\end{enumerate}
\end{enumerate}

\end{theorem}

\subsection{Suffix-Convex Languages}

{\bf LEFT IDEALS}
The complexity of left ideals was studied as follows:
complexities of common operations using various witnesses~\cite{BJL13},
semigroup size lower bound~\cite{BrYe11},  semigroup size upper bound~\cite{BrSz14a}, complexities of atoms~\cite{BrDa15}, 
most complex left ideals with restricted operations~\cite{BDL16},
most complex left ideals with restricted and unrestricted operations~\cite{BrSi16}.

\begin{definition}
\label{def:LWit}
For $n\ge 4$, let $\cD_n=\cD_n(a,b,c,d,e)=(Q_n,\Sig,\delta_n, 0, \{n-1\})$, where 
$\Sig=\{a,b,c,d,e\}$,
and $\delta_n$ is defined by  
transformations
$a\colon (1,\dots,n-1)$,
$b\colon(1,2)$,
${c\colon(n-1 \to 1)}$,
${d\colon(n-1\to 0)}$, 
and $e\colon (Q_n\to 1)$.
Denote by $L_n=L_n(a,b,c,d,e)$ the language accepted by~$\cD_n$.
\end{definition}

\begin{theorem}[Most Complex Left Ideals]
\label{thm:leftideals}
For each $n\ge 4$, the DFA of Definition~\ref{def:LWit} is minimal, and its language is a left ideal  of complexity $n$. 
The stream $(L_n(a,b,c,d,e) \mid n \ge 4)$  with some dialect streams
is most complex in the class of regular left ideals as follows:
1. Semigroup size: $n^{n-1}+n-1$.
2.  Quotient complexities: $n$.
3. Reversal: $2^{n-1}+1$.
4. Atom complexities:
 $\kappa(A_S) = n$, if $S=Q_n$; 
$\kappa(A_S) = 2^{n-1}$, if $S=\emp$;
$\kappa(A_S) = 1 + \sum_{x=1}^{|S|}\sum_{y=1}^{n-|S|}\binom{n-1}{x}\binom{n-x-1}{y-1}$, otherwise.
5. Star: $n+1$.
6. (a) Restricted product: $m+n+1$; (b) unrestricted product: $mn+m+n$.
7. Restricted and unrestricted boolean operations: same as regular languages.
  At least five letters are required to meet all these bounds~{\rm \cite{BSY15}}.
\end{theorem}

\noin
{\bf SUFFIX-CLOSED LANGUAGES}
The complexities of common operations using various witnesses were studied in~\cite{BJZ14}, and most complex suffix-closed languages in~\cite{BrSi17a}.

\begin{definition}
\label{def:sclosed}
For $n\ge 4$, let $\cD_n=\cD_n(a,b,c,d,e)=(Q_n,\Sig,\delta_n, 0, \{0\})$, where 
$\Sig=\{a,b,c,d,e\}$,
and $\delta_n$ is defined by  transformations
$a\colon (1,\dots,n-1)$,
$b\colon(1,2)$,
${c\colon(n-1 \to 1)}$,
${d\colon(n-1\to 0)}$, 
$e\colon (Q_n\to 1)$.
Let $L_n=L_n(a,b,c,d,e)$ be the language of~$\cD_n$.
\end{definition}

\begin{theorem}[Most Complex Suffix-Closed Languages]
\label{thm:sclosedmain}
For each $n\ge 4$, the DFA of Definition~\ref{def:sclosed} is minimal and its 
language $L_n(a,b,c,d,e)$ is suffix-closed and has complexity $n$.
The stream $(L_n(a,b,c,d,e) \mid n \ge 4)$  with some dialect streams
is most complex in the class of suffix-closed languages.
1. Semigroup size: $n^{n-1}+n-1$.
2. Quotients: $n$.
3. Reversal: $2^{n-1}+1$.
4. Atom complexities: 
 $ \kappa(A_S) = n$, if $S=\emp$;
$\kappa(A_S) = 2^{n-1}$, if $S=Q_n$;
$\kappa(A_S) = 1 + \sum_{x=1}^{|S|}\sum_{y=1}^{n-|S|}\binom{n-1}{y}\binom{n-y-1}{x-1}$, otherwise.
5. Star: $n$.
6. (a) Restricted product: $mn-n+1$; (b) unrestricted product: $mn+m+1$.
7. Restricted and unrestricted boolean operations: same as regular languages.
\end{theorem}

\noin
{\bf SUFFIX-FREE LANGUAGES}
The complexities of common operations using various witnesses were studied in~\cite{BrSi17a,BrSz17a,CmJi12,HaSa09,JiOl09}, 
semigroup size lower bound in~\cite{BLY12}, and upper bound in~\cite{BrSz17}.
 Suffix-free languages were the first example found of a class in which a most complex stream does not exist~\cite{BrSz17a}. 
However, two streams cover all the complexity measures~\cite{BrSz17a}. 
Since every suffix-free language has an empty quotient, the restricted and unrestricted cases for binary operations coincide.

\begin{definition}
\label{def:D6}
For $n\ge 4$, define the DFA 
$\mathcal{D}_n(a,b,c,d,e) =(Q_n,\Sigma,\delta,0,F),$
where $Q_n=\{0,\ldots,n-1\}$, $\Sigma=\{a,b,c,d,e\}$, 
 $\delta$ is given by 
$a\colon (0\to n-1) (1,\ldots,n-2)$,
$b\colon (0\to n-1) (1,2)$,
$c\colon (0\to n-1) (n-2\to 1)$,
$d\colon (\{0,1\}\to n-1)$,
$e\colon ( Q_n \setminus \{0\} \to n-1)(0\to 1)$,
and $F=\{q\in Q_n\setminus \{0,n-1\} \mid q \text{ is odd}\}$.
For $n=4$, $a$ and $b$ coincide, and we can use $\Sigma=\{b,c,d,e\}$.

Let $L_n(a,b,c,d,e)$ be the language of $\mathcal{D}_n(a,b,c,d,e)$.
\end{definition}

\begin{theorem}[Semigroup, Quotients, Reversal,  Atoms, Boolean Ops]
\label{thm:witness6} 
$L_n(a,b,c,d,e)$ is a suffix-free language of complexity $n$.
Moreover, it meets the following bounds:
1. Semigroup size: $(n-1)^{n-2}+n-2$ for $n\ge 6$.
2.  Quotient complexities: $n-1$, except $\kappa(L)=n$, $\kappa(\emp)=1$.
3. Reversal: $2^{n-2}+1$.
4. Atom complexities:
$ \kappa(A_S) = 2^{n-2}+1$, if $S=\emptyset$;
$ \kappa(A_S) = n$, if $S=\{0\}$;
$\kappa(A_S) = 1 + \sum_{x=1}^{|S|}\sum_{y=0}^{n-2-|S|}\binom{n-2}{x}\binom{n-2-x}{y}$, if $\emptyset\neq S\subseteq \{1,\ldots,n-2\}$.
5. Boolean operations: $mn - (m+n-2)$  if $\circ \in \{\cup, \oplus\}$, 
$mn - (m+2n-4)$ if $\circ = \setminus$, and 
$mn - 2(m+n-3)$ if $\circ = \cap$.

\end{theorem}

\begin{definition}
\label{def:suffree}
For $n\ge 4$, 
define the DFA 
$\mathcal{D}_n(a,b,c) =(Q_n,\Sigma,\delta,0,\{n-2\}),$
where $Q_n=\{0,\ldots,n-1\}$, $\Sigma=\{a,b,c\}$, 
and $\delta$ is defined by 
$ a \colon (0 \to n-1) (1,\dots,n-2) $,
$ b \colon (0 \to n-1) (1, 2)$, 
$ c \colon  (1,n-1) (0 \to 1) $.
Let $L_n(a,b,c)$ be the language of $\mathcal{D}_n(a,b,c)$.
\end{definition}

\begin{theorem}[Star, Product, Boolean Operations]
\label{thm:suffree}
$L_{n}(a,b,c)$ and its  dialects meet the bounds for star, product, and boolean operations as follows:
1. Star: $2^{n-2}+1$.
2. Product: $(m-1)2^{n-2}+1$.
3: Boolean operations: as in Theorem~\ref{thm:witness6}.
\end{theorem}

\noin 
{\bf  BIFIX-FREE LANGUAGES}

A language is \emph{bifix-free} if it is both prefix-free and suffix-free.
The complexities of common operations using various witnesses were studied in~\cite{BJLS14}, 
a conjecture on the semigroup size in~\cite{BLY12}, and tight upper bound in~\cite{SzWi16}. 
Since every bifix-free language has an empty quotient, the restricted and unrestricted cases for binary operations coincide.  The results below were found recently~\cite{FeSz16}.

\begin{definition}
\label{def:bifix-free_ops}
For $n\ge 7$, define the DFA 
$\mathcal{D}_n(a,b,c) =(Q_n,\Sigma,\delta,0,\{n-2\}),$
where $Q_n=\{0,\ldots,n-1\}$, $\Sigma=\{a,b,c\}$, 
$h= \lfloor (n-1)/2\rfloor$,
 $\delta$ is given by 
$a\colon (0\to 1) (\{1,\ldots,n-3\} \to n-2)(\{n-2,n-1\}\to n-1)$,
$b\colon (\{0,n-2,n-1\}\to n-1) (1,\ldots,n-3)$,
and $c\colon (\{0,n-2,n-1\}\to n-1) (1 \to h) (h \to n-2) ( n-3, \dots,h+1,h-1,\dots,2)$.
Let $L_n(a,b,c)$ be the language of $\mathcal{D}_n(a,b,c)$.
\end{definition}

\begin{theorem}[Bounds for Operations]
\label{thm:sfreemain}
The DFA of Definition~\ref{def:bifix-free_ops} is minimal and its 
language $L_n(a,b,c)$ is bifix-free and has complexity $n$.
The stream $(L_n(a,b,c) \mid n \ge 9)$  with some dialect streams
 meets the bounds for common operations on bifix-free languages:
 1. Quotient complexities: $n-1$, except $\kappa(L)=n$, 
$\kappa(\{\varepsilon\})=2$, and $\kappa(\emptyset)=1$.
2. Reversal: $2^{n-3}+2$.
3. Star: $n-1$.
4. Product: $m+n-2$.
5. Boolean operations: $mn - (m+n)$  if $\circ \in \{\cup, \oplus\}$, 
$mn - (2m+3n-9)$ if $\circ = \setminus$, and 
$mn - 3(m+n-4)$ if $\circ = \cap$.
\end{theorem}

Even though bifix-free languages are a subclass of suffix-free languages and there does not exist a most complex suffix-free stream, we do have a most complex bifix-free stream.
This stream has an alphabet of size 
$(n-2)^{n-3} + (n-3)2^{n-3} -1$~\cite{FeSz16}, and the alphabet size cannot be reduced.
The syntactic semigroup of this language is of size 
$(n-1)^{n-3} + (n-2)^{n-3} + (n-3) 2^{n-3}$. 
Maximal atom complexities are:
 $\kappa(A_S) = 2^{n-2}+1$, if $S=\emptyset$;
$\kappa(A_S) = n$, if $S=\{0\}$;
$\kappa(A_S) = 2$, if $S=\{n-2\}$; 
$\kappa(A_S) = 3 + \sum_{x=1}^{|S|}\sum_{y=0}^{n-3-|S|}\binom{n-3}{x}\binom{n-3-x}{y}$, if $\emptyset\neq S\subseteq \{1,\ldots,n-3\}$.

For further details see~\cite{FeSz16,SzWi16}.
\medskip

\noin
{\bf PROPER SUFFIX-CONVEX LANGUAGES}
This is the second class  found for which a most complex  stream  does not exist.
The complexity of this class is still being studied, but we do know that  at least three different witnesses are required to meet the bounds for all the measures\footnote{C. Sinnamon: private communication}.
\subsection{Factor-Convex Languages}

\noin
{\bf TWO-SIDED IDEALS} 
The complexities of basic operations on two-sided ideals were studied in~\cite{BJL13}.
The following stream of two-sided ideals was defined in~\cite{BrYe11}, where it was conjectured that the DFAs in this stream have maximal transition semigroups. This was proved in~\cite{BrSz14a},  and the stream was shown to be most complex for restricted operations in~\cite{BDL16}. It is also most complex in the unrestricted case~\cite{BrSi16}.

\begin{definition}
\label{def:2sided}
For $n\ge 5$, let  
$\cD_n =\cD_n(a,b,c,d,e,f)= (Q_n,\Sig,\delta, 0,\{n-1\})$, where
$\Sig=\{a,b,c,d,e,f\}$, 
and $\delta_n$ is defined by 
$a \colon (1,2,\ldots,n-2)$,
$b \colon (1,2)$,
$c \colon (n-2\to 1)$,
$d \colon (n-2\to 0)$,
$e \colon (Q_{n-1}\to 1)$,
and $f \colon (1 \to n-1)$.
Let $L_n(a,b,c,d,e,f)$ be the language of $\cD_n(a,b,c,d,e,f)$.
\end{definition}

\begin{theorem}[ Most Complex Two-Sided Ideals]
\label{thm:twosidedidealmain}
For $n\ge 5$, the language $L_n(a,b,c,d,e,f)$ is a two-sided ideal of complexity $n$.
The witness stream $(L_n(a,b,c,d,e,f) \mid n \ge 5)$  with some dialect streams
is most complex in the class of regular two-sided ideals,
and meets the following complexity bounds:
1. Semigroup size: $n^{n-2}+(n-2)2^{n-2}+1$. 
2.  Quotient complexities: $n$.
3. Reversal: $2^{n-1}+1$.
4. Atom complexities: 
 $\kappa(A_S) = n$, if $S=Q_n$; 
$\kappa(A_S) = 2^{n-2}+n-1$, if $S=Q_n\setminus \{1\} $; 
$\kappa(A_S) = 1 + \sum_{x=1}^{|S|}\sum_{y=1}^{n-|S|}\binom{n-2}{x-1}\binom{n-x-1}{y-1}$, otherwise.
5. Star: $n+1$. 
6. (a) Restricted product: $m+n-1$; 
(b) unrestricted product: $m+2n$.
7. (a) Restricted boolean operations: 
$mn$  if $\circ \in \{\cap, \oplus\}$, 
$mn - (m-1)$ if  $\circ=\setminus$,
$mn - (m+n-2)$ if $\circ=\cup$.
(b) unrestricted boolean operations: same as regular languages. 
  At least six letters are required to meet all these bounds~{\rm \cite{BrSi16}}.
\end{theorem}

\noin
{\bf FACTOR-CLOSED LANGUAGES}
The complexities of basic operations on factor-closed languages were examined in~\cite{BJZ14}. The syntactic complexity of factor-closed languages is the same as that of two-sided ideals, because each factor-closed language other than $\Sig^*$ is the complement of a two-sided ideal.
Most complex factor-closed languages have not been studied.
\medskip

\noin
{\bf FACTOR-FREE LANGUAGES}
The complexities of basic operations on factor-free languages were examined in~\cite{BJLS14}. The syntactic complexity of factor-free languages was conjectured in~\cite{BLY12} to be $(n-1)^{n-3}+ (n-3)2^{n-3}+1$, but the problem is still open.
Most complex factor-free languages have not been studied.
\medskip

\subsection{ Subword-Convex Languages}

\noin
{\bf ALL-SIDED IDEALS}
The complexities of basic operations  were examined in~\cite{BJL13}. The syntactic complexity has not been studied.
\medskip

\noin
{\bf SUBWORD-CLOSED LANGUAGES}
The complexities of basic operations  were examined in~\cite{BJZ14}. The syntactic complexity has not been studied.
\medskip

\noin
{\bf SUBWORD-FREE LANGUAGES}
The complexities of basic operations  were examined in~\cite{BJLS14}. The syntactic complexity has not been studied.

\subsection{Other Classes}

{\bf NON-RETURNING LANGUAGES}

A deterministic finite automaton (DFA) is \emph{non-returning} if there are no transitions into its initial state. 
A regular language is non-returning if its minimal DFA has that property. 
The state complexities of common operations (boolean operations, Kleene star, reverse and product) were studied by Eom, Han and Jir\'askov\'a~\cite{EHJ16}.
Most complex non-returning languages were examined in~\cite{BrDa17}.

If $t$ has rank $n-1$,  there is exactly one pair of distinct elements $i,j \in Q_n$ such that $it = jt$.
A transformation $t$ of $Q_n$ is of \emph{type $\{ i, j \}$} if $t$ has rank $n-1$ and $it = jt$ for $i < j$.

Let $\Gamma =\{a_{i,j} \mid 0\le i < j \le n-1\}$, where $a_{i,j}$ is a letter that induces any transformation of type $\{i,j\}$ and does not map any state to $0$. 
Let 
$\Gamma' = \Gamma \setminus \{ a_{0,n-1}, a_{0,1}, a_{1,n-1}, a_{0,2}\}$.
Let $\Sig = \{a,b,c,d\} \cup \Gamma'  $, where 
$a: (1,\dots,n-1) (0 \to 1)$,
$b:  (1,2) (0 \to 2)$, 
$c:  (2,\dotsc,n-1)(1 \to 2)(0 \to 1)$, and 
$d: (0 \to 2)$.

\begin{definition}
\label{def:most_complex}
For $n\ge 4$, let $\mathcal{D}_n=\mathcal{D}_n(\Sig)=(Q_n,\Sigma,\delta_n, 0, \{n-1\})$, where 
$\Sigma=\{a,b,c,d\} \cup \Gamma'$,
and $\delta_n$ is defined in accordance with the transformations described above.
Let $L_n=L_n(\Sig)$ be the language accepted by~$\mathcal{D}_n(\Sig)$.
\end{definition}

\begin{theorem}[Most Complex Non-Returning Languages]
\label{thm:nonreturning}
For each $n\ge 4$, the DFA of Definition~\ref{def:most_complex} is minimal and non-returning. 
The stream $(L_n(\Sig) \mid n \ge 4)$  with some dialect streams
is most complex in the class of regular non-returning languages and meets the bounds: 1. Semigroup size: $(n-1)^n$.  Quotient complexities: $n-1$, except $\kappa(L)=n$.
3. Reversal: $2^n$. 4. Atom complexities:
$ \kappa(A_S) = 2^{n-1}$, if $S\in \{\emp,Q_n\}$;
$ \kappa(A_S) = 2+ \sum_{x=1}^{|S|} \sum_{y=1}^{|S|} \binom{n-1}{x}\binom{n-1-x}{y}$, otherwise.
5. Star: $2^{n-1}$.
6. (a) Restricted product: $(m-1)2^{n-1}+1$;
(b) unrestricted product: $m2^{n-1}+1$.
7. (a) Restricted boolean operations: $mn-(m+n-2)$;
(b) unrestricted boolean operations:
	$(m+1)(n+1)$ if $\circ\in \{\cup,\oplus\}$,
 	$mn+m$ if  $\circ=\setminus $,
 	 $mn$ if $\circ= \cap$. 
	 
The bound on the semigroup size and on the complexity of atoms require an alphabet of at least $\binom{n}{2}$ letters, reversal requires at least three letters, and all the other bounds can be met by binary witnesses.
\end{theorem}

\goodbreak

\noin
{\bf STAR-FREE LANGUAGES}
 
 A language is \emph{star-free} if it 
 can be constructed from the basic languages using only boolean operations  and product, but no star.
 A famous theorem by Sch\"utzenberger~\cite{Sch65} states that a language is star-free if and only if its syntactic  monoid is aperiodic, meaning that it contains only trivial one-element groups. Star-free languages have  many interesting subclasses~\cite{Brz76}.
 
 The complexities of basic operations on star-free languages were studied in~\cite{BrLiu12}. 
 It is surprising that these languages can meet all the bounds for regular languages, except for reversal, which has a tight upper bound of $2^n-1$~\cite{BrSz14}. Most complex star-free languages have not been studied mainly because no tight upper bound on  their syntactic complexity is known, even though some large aperiodic semigroups have been found~\cite{BrSz15}.

Syntactic complexities for several subclasses of star-free languages have been found:
\be
\item
A language is $J$-trivial if its syntactic monoid $M$ satisfies  the following:$MsM=MtM$ implies $s=t$, for all $s,t \in M$.
It has been shown in~\cite{BrLi14} that the syntactic complexity of $J$-trivial  languages is $\lfloor e(n-1)! \rfloor$.
\item
A language is $R$-trivial if its syntactic monoid $M$ satisfies  the following: $sM=tM$ implies $s=t$, for all $s,t \in M$.
The syntactic complexity of $R$-trivial languages is $n!$~\cite{BrLi14}.
\item
A language is \emph{cofinite} if its complement is finite. The syntactic complexity of the class of finite and cofinite languages  is $(n-1)!$~\cite{BLL12}.
\item
A language is \emph{reverse definite} if it can be expressed in the form 
$L=E\cup F\Sig^*$, where $E$ and $F$ are finite. The syntactic complexity of reverse definite languages  is $(n-1)!$~\cite{BLL12}.
\ee

\section{Groups and Complexity}
\label{sec:groups}
We close this paper with a brief mention of some  group-theoretic results that simplify certain proofs about complexity. 

Let $S_n$ denote the symmetric group of degree $n$. 
A \emph{basis}
of $S_n$
is an ordered pair $(s,t)$ of distinct transformations of $Q_n=\{0,\dots,n-1\}$ that generate $S_n$.
Two bases $(s,t)$ and $(s',t')$ of $S_n$ are \emph{conjugate} if there exists a transformation $r\in S_n$ such that $rsr^{-1}=s'$, and  $rtr^{-1}=t'$.

Assume that a DFA $\cD'_m$ (respectively, $\cD_n$) has state set $Q'_m$ ($Q_n$), and let the subgroup of permutations of its
transition semigroup be $S_m$ ($S_n$).
Let $L'_m$ ($L_n$) be the language accepted by $\cD'_m$ ($\cD_n$).
The following was proved in~\cite{BBMR14}:

\begin{theorem}
\label{thm:BBMR}
Suppose $m,n\ge 2$ and $(m,n)\not \in \{(2,2),(3,4),(4,3),(4,4)\}$.
If   the subgroups of permutations in  the transition semigroups of $\cD'_m$ and $\cD_n$ are $S_m$ and $S_n$ respectively, and $\circ$ is a proper binary boolean operation,
then the complexity of $L'_m\circ L_n$ is $mn$, unless
$m=n$ and the bases induced by the letters of $\Sig$  in the transition semigroups of $\cD'_m$ and $\cD_n$ are conjugate, in which case the quotient complexity of $L'_m\circ L_n$ is at most $m=n$.
\end{theorem}
 In the DFAs used for most complex streams, usually the transition semigroups contain all permutations of some subset of the state set.
Theorem~\ref{thm:BBMR} has greatly simplified the proofs of results about the complexity of boolean operations in several cases~\cite{BrDa17,BDL16,BrSi16,BrSi17,BrSz14a}.
\medskip

 In the special case where $\cD'_m$ and $\cD_n$ are DFAs with exactly one final state, which occurs very commonly in most complex streams, there is a stronger result due to Davies~\cite{Dav17}.

Recall that to recognize boolean operations on the languages of $\cD'_m$ and $\cD_n$, we use the direct product DFA $\cD'_m \times \cD_n$ with state set $Q'_m \times Q_n$. A \emph{row} of $Q'_m \times Q_n$ is a set of the form $R_{p'} = \{(p',q) : q \in Q_n\}$. A \emph{column} of $Q'_m \times Q_n$ is a set of the form $C_{q} = \{(p',q) : p' \in Q'_m\}$.

We say a state $q$ of a DFA $\cD$ is \emph{reachable by permutations} if it is reachable by some word $w$ that induces a permutation in the transition semigroup of $\cD$. If the transition semigroup of $\cD$ is a group, this is the same thing as just being reachable.

\begin{theorem}
Suppose $m,n \ge 2$ and $(m,n) \ne (2,2)$.
Let $\cD'_m$ and $\cD_n$ be minimal DFAs with $m$ and $n$ states respectively.
Suppose that every state in each DFA is reachable by permutations, and each DFA has exactly one final state.
Consider the direct product $\cD'_m \times \cD_n$.
The following are equivalent:
\be
\item
Every state in $Q'_m \times Q_n$ is reachable by permutations.
\item
There exists $p' \in Q'_m$ such that every state in row $R_{p'} \subseteq Q'_m \times Q_n$ is reachable by permutations.
\item
There exists $q \in Q_n$ such that every state in column $C_q \subseteq Q'_m \times Q_n$ is reachable by permutations.
\item
The complexity of $L'_m \circ L_n$ is $mn$ for all proper binary boolean operations $\circ$.
\ee
\end{theorem}
It was proved in~\cite{BBMR14} that if $\cD'_m$ and $\cD_n$ satisfy the conditions of Theorem~\ref{thm:BBMR}, then every state in $Q'_m \times Q_n$ is reachable by permutations.
However, this result applies more generally, including in cases where the transition semigroups of $\cD'_m$ and $\cD_n$ are not symmetric groups.
The downside is the restriction on the final state sets.
\medskip

 The next result\footnote{S. Davies: private communication} greatly simplified a proof about product~\cite{BrSi16}.
Suppose $\cD'_m=( Q'_m, \Sig, \delta', 0', \{f'\} )$ is a minimal DFA of $L'_m$, $f'\neq 0'$, and 
$\cD_n=(Q_n,\Sig,\delta,0,F) $ is a minimal DFA of  $L_n$.
 We use the normal construction of an $\eps$-NFA $\cN$ -- an NFA that permits also transitions induced by the empty word -- to recognize $L'_mL_n$, by introducing an $\eps$-transition from the final state of $\cD'_m$ to the initial state of $\cD_n$, and changing the final state of $\cD'_m$ to non-final. 
We need to show that the following types of sets are reachable from the initial set $\{0'\}$ in the subset construction for $\cN$: (a)  
$(m-1)2^n$ sets $\{p'\} \cup S$, where $p'\in Q'_m\setminus \{f'\}$, and $S\subseteq Q_n$, 
(b) $2^{n-1}$ sets  
$\{ f' , 0\} \cup S$, where  $S\subseteq Q_n\setminus \{0\}$.
The lemma below allows us to check the reachability of only a few special sets.

\begin{lemma}
If
the transition semigroups of $\cD'_m$ and $\cD_n$ are groups,
and all the sets of the form
$ \{p'\},\; p'\in Q_m' \setminus \{f'\},  \text{ and  } \{0',q\}, \; q \in Q_n $
are reachable, then so are all sets of the form 
$$ \{p'\} \cup S,\; p'\in Q_m' \setminus \{f'\}, 
\; S\subseteq Q_n \text{ and  } \{f',0\} \cup S, \; S \subseteq Q_n\setminus \{0\}.  $$
\end{lemma} 

\section{Conclusions}
We have surveyed many papers concerned with complexity measures for regular languages and finite automata, and put special emphasis on most complex languages because they concisely describe the properties of the given languages. However, some questions remain. 

 Upper bounds on syntactic complexity are known for several subclasses.
Finding these upper bounds was trivial for right ideals and non-returning languages, easy for prefix-free and proper prefix-convex languages, and challenging for left ideals and two-sided ideals, suffix-free and bifix-free languages. 
The problem remains open for factor-free, and subword-free languages, and all-sided ideals.
As well, this question is open for star-free languages and many proper subclasses of star-free languages~\cite{Brz76}, for example, definite languages~\cite{BLL12} and $L$-trivial languages, where a language is $L$-trivial if its syntactic monoid $M$ satisfies  the following: $Ms=Mt$ implies $s=t$, for all $s,t \in M$.
Of course, if the syntactic complexity is unknown, then so is the existence of most complex language streams.
 
We have included atom complexities as a measure, but more work needs to be done to determine their usefulness. We justify the inclusion of atom complexities by the following observation: so far, whenever a language stream meets the bounds for the basic operations and syntactic complexity, it also meets the bounds for atom complexities. 

Finally, it would be very useful to have more results like those in Section~\ref{sec:groups} because they allow us to avoid complex proofs or greatly simplify them.
\medskip

\noin
{\bf Acknowledgment}
I am very grateful to Sylvie Davies, Corwin Sinnamon, Marek Szyku{\l}a, and Hellis Tamm not only for careful proofreading, but also for their many contributions to this paper.


\providecommand{\noopsort}[1]{}

\end{document}